\documentclass[fleqn,12pt,twoside]{article}
\usepackage{amsmath}
\usepackage{espcrc1}
\usepackage{graphicx}
\usepackage[figuresright]{rotating}

\newcommand{\Eq}{Eq.}
\newcommand{\Eqs}{Eqs.}
\newcommand{\Ref}{Ref.}
\newcommand{\Refs}{Refs.}
\newcommand{\mbf}[1]{{\mathbf{#1}}}
\newcommand{\fm}{\;\mathrm{fm}}

\newcommand{\cm}{\mathrm{c\!\:\!.m\!\:\!.}}
\newcommand{\zr}{z_R^{-\frac12}}
\newcommand{\ZaR}{\mathcal{Z}_{\alpha R}^{-\frac12}}
\newcommand{\ZbR}{\mathcal{Z}_{\beta R}^{-\frac12}}
\newcommand{\zR}{\mathcal{Z}}

\title{Treatment of the Coulomb interaction in three-nucleon reactions}

\author{A. Deltuva\address[CFNUL]{Centro de F\'{\i}sica Nuclear 
    da Universidade de Lisboa, P-1649-003 Lisboa, Portugal}\thanks
  {Supported by the FCT grant SFRH/BPD/14801/2003},
  A. C. Fonseca\addressmark[CFNUL]
  and
  P. U. Sauer\address{Institut f\"ur Theoretische Physik,  
    Universit\"at Hannover,  D-30167 Hannover, Germany} }

\begin{document}

\maketitle

\begin{abstract}
The Coulomb interaction between the two protons is included in the calculation
of three-nucleon hadronic and electromagnetic reactions using
screening and renormalization approach.
Calculations are done using integral equations in momentum space.
The reliability of the method is demonstrated.
The Coulomb effect on observables is discussed.
\end{abstract}

\section{Introduction \label{sec:intro}}

The inclusion of the Coulomb interaction in the description of the
three-nucleon continuum is one of the most challenging tasks in
theoretical few-body nuclear physics.
The Coulomb interaction is well known, in contrast to the strong
two-nucleon and three-nucleon potentials mainly studied in three-nucleon
scattering. However, due to its $1/r$ behavior, the Coulomb interaction does
not satisfy the mathematical properties required for the formulation
of standard scattering theory.
There is a long history of theoretical work on the solution of the
Coulomb problem in three-particle scattering. Some of the recent 
suggestions \cite{alt:04a,kadyrov:05a,oryu:06a}  have not matured yet
into practical applications, while the others, based mostly on the
configuration-space framework  
\cite{chen:01a,ishikawa:03a,doleschall:05a}, are limited to energies below
deuteron breakup threshold (DBT). Up to now only few approaches led
to the results above DBT. Those are configuration-space calculations
for proton-deuteron $(pd)$ elastic scattering 
using the Kohn variational principle \cite{kievsky:01a} and
the screening and renormalization approach in the framework of
momentum-space integral equations 
\cite{alt:78a,alt:94a,alt:02a,deltuva:05a,deltuva:05d}; nevertheless,
for the latter method only the present work published in
\Refs~\cite{deltuva:05a,deltuva:05d} uses realistic interactions together
with fully converged calculations in terms of screening radius and
two-nucleon and three-nucleon partial waves.

Section~\ref{sec:th} shortly recalls the technical apparatus 
underlying the calculations and demonstrates the reliability of the method.
Section~\ref{sec:summ} gives our summary.

\section{Screening and renormalization approach \label{sec:th} } 

Our treatment of the Coulomb interaction is based on the idea of
screening and renormalization proposed in \Ref~\cite{taylor:74a}
for the scattering of two charged particles.
The screened Coulomb potential of our choice in $r$-space representation
is given by
\begin{gather} \label{eq:wr}
w_R(r) = w(r)\; e^{-(r/R)^n},
\end{gather}
where $w(r)$ is the proper Coulomb potential, $R$ is the screening radius,
and $n$ controls the smoothness of the screening.
The standard scattering theory is formally applicable to the screened 
Coulomb potential, e.g., the transition matrix is defined via 
Lippmann-Schwinger equation
$t_R(e_i+i0) = w_R + w_R g_0(e_i+i0) t_R(e_i+i0)$ 
and the corresponding wave function
is $|\psi_R^{(+)}(\mbf{p}_i) \rangle = 
[1 + g_0(e_i+i0) t_R(e_i+i0)] |\mbf{p}_i \rangle $,
where $g_0(e_i+i0)$ is the free resolvent and $|\mbf{p}_i \rangle$ is 
the plane-wave state with momentum 
$\mbf{p}_i$ and energy $e_i$.
Furthermore, as shown in \Refs~\cite{taylor:74a,gorshkov:61},
the  on-shell screened Coulomb transition matrix 
$\langle \mbf{p}_f| t_R (e_i+i0) |\mbf{p}_i \rangle$ with  $p_f=p_i$
and wave function diverge in $R \to \infty$ limit, but after renormalization 
with also diverging phase factor $z_R(p_i)$
they converge to the proper Coulomb amplitude 
$\langle \mbf{p}_f| t_C |\mbf{p}_i \rangle$
and proper Coulomb wave function $|\psi_C^{(+)}(\mbf{p}_i) \rangle $, 
respectively:
\begin{subequations} \label{eq:SR}
  \begin{gather} \label{eq:tc}
    \lim_{R \to \infty}
    \langle \mbf{p}_f| t_R (e_i+i0) |\mbf{p}_i \rangle z_R^{-1}(p_i) 
    =  \langle \mbf{p}_f| t_C |\mbf{p}_i \rangle, \\
    \lim_{R \to \infty}
    |\psi_R^{(+)}(\mbf{p}_i) \rangle  z_R^{-\frac12}(p_i) 
    =  |\psi_C^{(+)}(\mbf{p}_i) \rangle.
  \end{gather}
\end{subequations}
The renormalized screened Coulomb amplitude converges to 
the proper Coulomb amplitude in general as a distribution.
As discussed in \Ref~\cite{taylor:74a}, this is fully
sufficient for description of real experiments and justifies
the replacement of $\lim_{R \to \infty}
    \langle \mbf{p}_f| t_R (e_i+i0) |\mbf{p}_i \rangle z_R^{-1}(p_i) $
by $ \langle \mbf{p}_f| t_C |\mbf{p}_i \rangle$ in practical calculations
\cite{alt:02a,deltuva:05a}.

The screening and renormalization approach can be applied to
more complicated systems, if in the scattering amplitudes
the diverging screened Coulomb contributions can be isolated
in the form of two-body on-shell transition matrix
and two-body wave function with known renormalization properties
\eqref{eq:SR}. For the description of $(pd)$ scattering
we employ Alt-Grassberger-Sandhas (AGS) three-particle scattering
equations~\cite{alt:67a} in momentum space
\begin{subequations}\label{eq:AGS}
  \begin{gather} \label{eq:Uba}
     U^{(R)}_{\beta \alpha}(Z) =  \bar{\delta}_{\beta \alpha} G_0^{-1}(Z)
     + \sum_{\sigma} \bar{\delta}_{\beta \sigma} T^{(R)}_\sigma (Z) G_0(Z)
     U^{(R)}_{\sigma \alpha}(Z), \\
     U^{(R)}_{0 \alpha}(Z) =  G_0^{-1}(Z)
     + \sum_{\sigma}  T^{(R)}_\sigma (Z) G_0(Z) U^{(R)}_{\sigma \alpha}(Z),
  \end{gather}
\end{subequations}
with  $\bar{\delta}_{\beta \alpha} = 1 - {\delta}_{\beta \alpha}$,
$ G_0(Z)$ being the free resolvent, 
$T^{(R)}_\sigma (Z)$ the two-particle transition matrix
derived from nuclear plus screened Coulomb potentials, and 
$U^{(R)}_{\beta \alpha}(Z)$  and $U^{(R)}_{0 \alpha}(Z)$ 
the three-particle transition operators
for elastic/rearrangement and breakup scattering; their dependence on the 
screening radius $R$ is notationally indicated.
As demonstrated in \Refs~\cite{alt:78a,deltuva:05a,deltuva:05d},
the three-particle transition operators can be decomposed into
long-range and Coulomb-distorted short-range parts 
\begin{subequations}\label{eq:U-T}
\begin{gather} 
    U^{(R)}_{\beta \alpha}(Z) = 
    \delta_{\beta\alpha} T^{\cm}_{\alpha R}(Z)
    +  [1 + T^{\cm}_{\beta R}(Z) G^{(R)}_{\beta}(Z)] 
    \tilde{U}^{(R)}_{\beta\alpha}(Z)
          [1 + G^{(R)}_{\alpha}(Z) T^{\cm}_{\alpha R}(Z)], \\
      U^{(R)}_{0\alpha}(Z) = [1 + T_{\rho R}(Z) G_{0}(Z)]
      \tilde{U}^{(R)}_{0\alpha}(Z) 
            [1 + G^{(R)}_{\alpha}(Z) T^{\cm}_{\alpha R}(Z)]
\end{gather}
\end{subequations}
with the channel resolvent $G^{(R)}_{\alpha}(Z)$,
proton-proton $(pp)$ screened Coulomb transition matrix $T_{\rho R}(Z)$, 
two-body transition matrix $T^{\cm}_{\alpha R}(Z) $
derived from the screened Coulomb potential between
spectator and the center of mass (c.m.) of the remaining pair,
and with the reduced short-range operators $\tilde{U}^{(R)}_{\beta\alpha}(Z)$
and $\tilde{U}^{(R)}_{0\alpha}(Z)$.
On-shell matrix elements of the operators \eqref{eq:U-T}
 between  two- and- three-body
channel states  $|\phi_\alpha (\mbf{q}_i) \nu_{\alpha_i} \rangle $
and $|\phi_0 (\mbf{p}_f \mbf{q}_f) \nu_{0_f} \rangle $ with 
discrete quantum numbers $\nu_{\sigma_j} $,
Jacobi momenta $\mbf{p}_j $ and  $\mbf{q}_j $,  and energy  $E_i$,
do not have a $R \to \infty$ limit. However, the quantities diverging 
in that limit are already isolated in \Eqs~\eqref{eq:U-T} and are
of two-body nature, i.e., the on-shell
$T^{\cm}_{\alpha R}(Z)$ and the initial/final state screened Coulomb
wave functions. Those quantities, renormalized according to
\Eq~\eqref{eq:SR}, in the $R \to \infty$ limit converge to 
the two-body Coulomb scattering amplitude $T^{\cm}_{\alpha R} $ 
and to the corresponding Coulomb wave functions, respectively,
thereby yielding the $pd$ scattering amplitudes
in the proper Coulomb limit
\begin{subequations}\label{eq:UC}
\begin{gather} \label{eq:UC1}
  \begin{split}
    \langle  \phi_\beta (\mbf{q}_f)  \nu_{\beta_f} | U_{\beta \alpha}
    |\phi_\alpha (\mbf{q}_i) \nu_{\alpha_i} \rangle   = {}&
    \delta_{\beta \alpha}
    \langle \phi_\alpha (\mbf{q}_f) \nu_{\alpha_f} |T^{\cm}_{\alpha C}
    |\phi_\alpha (\mbf{q}_i) \nu_{\alpha_i} \rangle  \\ & +
    \lim_{R \to \infty} \{ \ZbR(q_f)
    \langle \phi_\beta (\mbf{q}_f) \nu_{\beta_f} |
            [ U^{(R)}_{\beta \alpha}(E_i + i0)  \\ & -
              \delta_{\beta\alpha} T^{\cm}_{\alpha R}(E_i + i0)]
            |\phi_\alpha (\mbf{q}_i) \nu_{\alpha_i} \rangle
            \ZaR(q_i) \},
  \end{split} \\
  \begin{split}
      \langle \phi_0  (\mbf{p}_f \mbf{q}_f) \nu_{0_f} | U_{0 \alpha}
      |\phi_\alpha (\mbf{q}_i) \nu_{\alpha_i} \rangle  = {} &
      \lim_{R \to \infty} \{ \zr(p_f)
      \langle \phi_0 (\mbf{p}_f \mbf{q}_f) \nu_{0_f} | \\ &  \times
      U^{(R)}_{0 \alpha}(E_i + i0)
      |\phi_\alpha (\mbf{q}_i) \nu_{\alpha_i} \rangle \ZaR(q_i) \},
    \end{split}
  \end{gather}
\end{subequations}
where the relation between \Eqs~\eqref{eq:UC} and \eqref{eq:U-T} is
given in \Refs~\cite{deltuva:05a,deltuva:05d}.
The renormalization factors $\zR_{\alpha R}$ and $z_R$ are diverging phase 
factors defined in \Ref~\cite{taylor:74a,deltuva:05a,deltuva:05d}.
The $R \to \infty$ limit in \Eqs~\eqref{eq:UC}
has to be calculated numerically, but due to the short-range nature
of the corresponding operators it can be reached with sufficient accuracy 
at rather modest $R$ if the form of the screened Coulomb potential
\eqref{eq:wr}, in particular the parameter $n$ controlling the smoothness 
of the screening, has been chosen successfully.
We prefer to work with a sharper screening than the Yukawa screening
$(n=1)$ of \Refs~\cite{alt:94a,alt:02a}. We want to ensure that the
screened Coulomb potential $w_R(r)$ approximates well the true Coulomb one
$w(r)$ for distances $r<R$  and simultaneously vanishes rapidly for $r>R$,
providing a comparatively fast convergence of the partial-wave expansion.
In contrast, the sharp cutoff  $(n \to \infty)$
yields an unpleasant oscillatory behavior in the momentum-space representation,
leading to convergence problems. In \Refs~\cite{deltuva:05a,deltuva:05d}
we found the values $3 \le n \le 6$ to provide a sufficiently smooth,
but at the same time a sufficiently rapid screening around $r=R$;
$n=4$ is our standard choice.
The screening radius $R$ sufficient for convergence in \Eqs~\eqref{eq:UC}
is considerably larger than the range of the strong interaction.
As a consequence, the calculation of the three-particle transition
operators for nuclear plus screened Coulomb potentials requires
the inclusion of partial waves with angular momentum much higher
than required for the hadronic potential alone.
This problem can be solved in efficient and reliable way either by using
the perturbative approach for high two-particle partial waves, developed in
\Ref~\cite{deltuva:03b}, or even without it as discussed in 
\Ref~\cite{deltuva:06a}.
More details on the practical implementation of the screening
and renormalization approach as well as the extension to three-nucleon
electromagnetic (e.m.) reactions are presented in 
\Refs~\cite{deltuva:05a,deltuva:05d}.

\begin{figure}[t]
\includegraphics[scale=0.66]{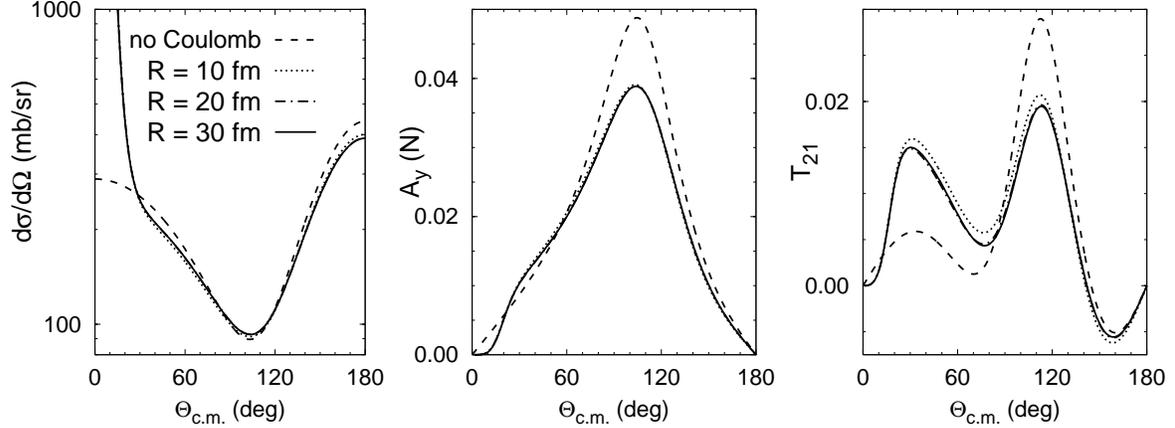}\vspace{-8mm}
\caption{\label{fig:Rconv1}
Convergence of the $pd$ elastic scattering  observables with screening radius 
$R$. The differential cross section, proton analyzing power $A_y(N)$ and
deuteron analyzing power $T_{21}$ at 3~MeV proton lab energy  are
shown as functions of the c.m. scattering angle.
Results obtained with screening radius $R= 10$~fm (dotted curves), 
20~fm (dash-dotted curves), and 30~fm (solid curves) are compared. 
Results without Coulomb (dashed curves)
are given as reference for the size of the Coulomb effect.}
\end{figure}

\begin{figure}[!]
\includegraphics[scale=0.61]{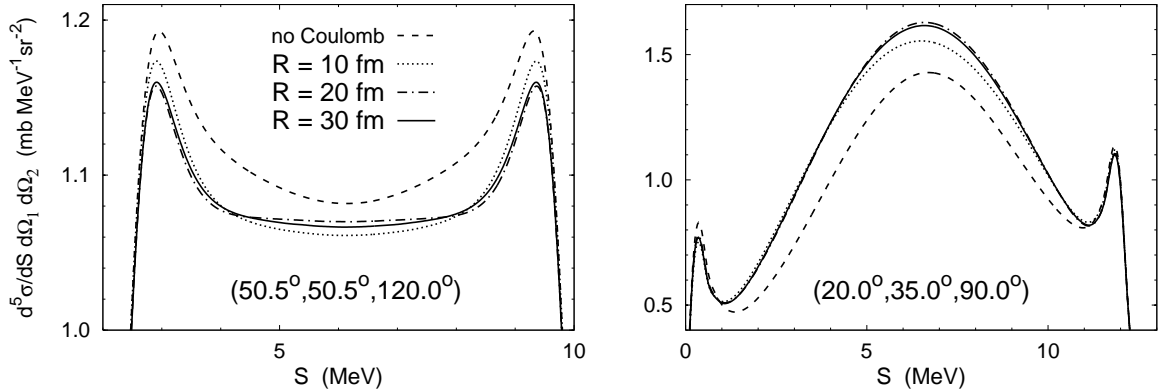}\vspace{-8mm}
\caption{\label{fig:Rconv2}
Convergence of the $pd$ breakup  observables with screening radius $R$.
The differential cross section in selected kinematical configurations at
13~MeV proton lab energy is shown as function of the arclength $S$ 
along the kinematical curve. Notation of curves as in Fig.~\ref{fig:Rconv1}.}
\end{figure}

\begin{figure}[!]
\begin{center}
\includegraphics[scale=0.6]{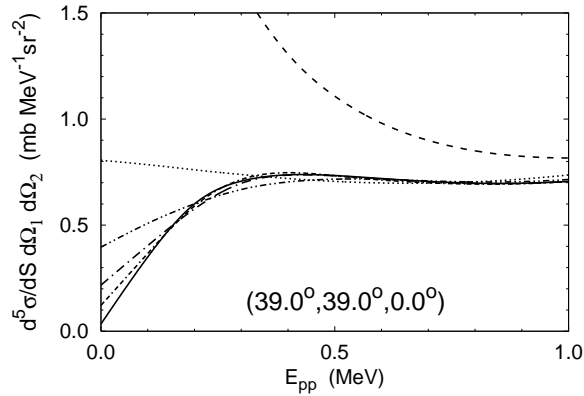}
\end{center} \vspace{-11mm}
\caption{\label{fig:Rconv3}
Convergence of the $pd$ breakup  observables with screening radius $R$.
The differential cross section  for $pd$ breakup at 13~MeV
proton lab energy in the $pp$-FSI configuration is shown
as function the relative $pp$ energy $E_{pp}$.
Results obtained with screening radius $R= 10$~fm (dotted curve),
20~fm (dashed-double-dotted curve),
30~fm (dashed-dotted curve), 40~fm (double-dashed-dotted curve), 
60~fm (solid curve), and results without Coulomb (dashed curve)
are compared.}
\end{figure}

The internal criterion for the reliability of our method is the convergence
of the observables with screening radius $R$ employed to calculate the 
Coulomb-distorted short-range part of the amplitudes in \Eqs~\eqref{eq:UC}.
Figures \ref{fig:Rconv1} and \ref{fig:Rconv2}
show characteristic examples for $pd$ elastic scattering and breakup.
The hadronic interaction is the realistic coupled-channel potential
CD Bonn + $\Delta$, allowing for single virtual $\Delta$-isobar excitation
\cite{deltuva:03c}. 
In most cases the convergence is impressively fast;
the screening radius $R = 20 \fm$ is sufficient.
The exceptions requiring larger screening radii are the $pd$ elastic
scattering observables at very low energies and the breakup
differential cross section in kinematical
situations characterized by very low $pp$ relative energy $E_{pp}$, i.e.,
close to the $pp$ final-state interaction  ($pp$-FSI) regime,
as shown in  Fig.~\ref{fig:Rconv3}.
In there, the $pp$ repulsion is responsible for decreasing the cross section,
converting the $pp$-FSI peak obtained in the absence of Coulomb
into a minimum with zero cross section at $p_f=0$, i.e., for $E_{pp}=0$.
Such a behavior is seen in the experimental data as well 
\cite{kistryn:06a,sagara:fb18}.
The slow convergence close to $pp$-FSI is not surprising, since the 
renormalization factor $z_R(p_f=0)$ itself is ill-defined,
indicating that the  screening and renormalization
procedure cannot be applied at $p_f=0$.
Therefore an extrapolation has to be used to calculate
the observables at $p_f=0$, which works pretty well since
the observables vary smoothly with $p_f$. 

Furthermore, \Ref~\cite{deltuva:05b} makes a detailed comparison between the
results  for $pd$ elastic scattering obtained by the present technique and 
those of \Ref~\cite{kievsky:01a} obtained from the variational solution
of the three-nucleon Schr\"odinger equation in configuration space
with the inclusion of an \emph{unscreened} Coulomb potential
between the protons and imposing  the proper Coulomb boundary
conditions explicitly. The agreement, across the board, between the
results derived from two entirely different methods, clearly
indicates that both techniques for including the Coulomb interaction
are reliable.

\section{Summary \label{sec:summ}}

We have shown how the Coulomb interaction between the protons
can be included into the momentum-space description of proton-deuteron 
scattering
using old idea of screening and renormalization \cite{taylor:74a}.
The theoretical framework is the AGS integral equation \cite{alt:67a}.
The calculations are done on the same level of accuracy and
sophistication as for the neutron-deuteron scattering.
Our practical realization  differs enormously 
from the one of \Refs~\cite{alt:94a,alt:02a}, even the form of three-particle
equations including screened Coulomb is  different. 
We use modern hadronic interactions whereas the calculations 
of \Refs~\cite{alt:94a,alt:02a} were based on quasiparticle equations with 
rank-1 separable potentials and in addition approximated the screened Coulomb 
transition matrix by the screened Coulomb potential.

Compared to the configuration-space treatment \cite{kievsky:01a}, 
the results for the elastic $pd$ scattering agree
very well over a wide energy range \cite{deltuva:05b}.
Although at very low energies
the coordinate-space methods remain favored, at higher energies and 
especially for three-body breakup reactions our momentum-space treatment 
is more efficient, and so far the only one to show first results
for realistic interactions \cite{deltuva:05c}.

A realm of applications to the rich $pd$ data base for elastic scattering
and breakup may be seen in 
\Refs~\cite{deltuva:05a,deltuva:05d,kistryn:06a,sagara:fb18}.
The conclusion is that in elastic $pd$ scattering the Coulomb effect  is 
important at low energies for all kinematic regimes, but gets confined 
to the forward direction at higher energies.
In $pd$ breakup and in three-body e.m. disintegration of ${}^3\mathrm{He}$ 
the Coulomb effect is extremely important in kinematical regimes close to 
$pp$-FSI. There the $pp$ repulsion converts the $pp$-FSI peak obtained in the
absence of Coulomb into a minimum with zero cross section \cite{sagara:fb18}.
This significant change of the cross section behavior has important
consequences in nearby configurations where one may observe
instead an increase of the cross section due to Coulomb \cite{kistryn:06a}.
However, some of the long-standing discrepancies between experiment and theory
like the space star anomaly in $pd$ breakup are not resolved by the inclusion
of the Coulomb interaction.

Finally, the screening and renormalization approach for including the Coulomb 
interaction is  extended  to  four-nucleon scattering \cite{fonseca:fb18}.


\begin{thebibliography}{10}

\bibitem{alt:04a}
E.~O. Alt, S.~B. Levin, S.~L. Yakovlev, Phys.~Rev.~C { 69} (2004) 034002.

\bibitem{kadyrov:05a}
A.~S. Kadyrov, I. Bray, A.~M. Mukhamedzhanov, A.~T. Stelbovics,
  Phys.~Rev.~A { 72} (2005)  032712 .

\bibitem{oryu:06a}
S. Oryu, Phys.~Rev.~C { 73}  (2006)  054001.

\bibitem{chen:01a}
C.~R. Chen, J.~L. Friar, G.~L. Payne, Few-Body Syst. { 31}  (2001)  13.

\bibitem{ishikawa:03a}
S. Ishikawa, Few-Body Syst. { 32}  (2003)  229.

\bibitem{doleschall:05a}
P. Doleschall and Z. Papp, Phys.~Rev.~C { 72}  (2005)  044003.

\bibitem{kievsky:01a}
A. Kievsky, M. Viviani, S. Rosati, Phys.~Rev.~C { 64}  (2001)  024002.

\bibitem{alt:78a}
E.~O. Alt, W. Sandhas, H. Ziegelmann, Phys.~Rev.~C { 17}  (1978)  1981;
E.~O. Alt and W. Sandhas, {\it ibid.} { 21} (1980) 1733.

\bibitem{alt:94a}
E.~O. Alt and M. Rauh, Few-Body Syst. { 17}  (1994)  121.

\bibitem{alt:02a}
E.~O. Alt, A.~M. Mukhamedzhanov, M.~M. Nishonov, A.~I. Sattarov,
  Phys.~Rev.~C { 65}  (2002)  064613.

\bibitem{deltuva:05a}
A. Deltuva, A.~C. Fonseca, P.~U. Sauer, Phys.~Rev.~C { 71}  (2005)  054005.

\bibitem{deltuva:05d}
A. Deltuva, A.~C. Fonseca, P.~U. Sauer, Phys.~Rev.~C { 72}  (2005)  054004.

\bibitem{taylor:74a}
J.~R. Taylor, Nuovo Cimento { B23}  (1974)  313; M.~D. Semon and J.~R.
  Taylor, {\it ibid.} { A26} (1975) 48.

\bibitem{gorshkov:61}
V.~G. Gorshkov, Sov.~Phys.-JETP { 13}  (1961)  1037.

\bibitem{alt:67a}
E.~O. Alt, P. Grassberger, W. Sandhas, Nucl.~Phys. { B2}  (1967)  167.

\bibitem{deltuva:03b}
A. Deltuva, K. Chmielewski, P.~U. Sauer, Phys.~Rev.~C { 67}, (2003) 054004.

\bibitem{deltuva:06a}
A. Deltuva, A.~C. Fonseca, P.~U. Sauer, Phys.~Rev.~C { 73}  (2006)  057001.

\bibitem{deltuva:03c}
A. Deltuva, R. Machleidt,  P.~U. Sauer, Phys.~Rev.~C { 68}  (2003)  024005.

\bibitem{kistryn:06a}
S. Kistryn~{\it et al.}, Phys.~Lett.~B  641  (2006)  23.

\bibitem{sagara:fb18}
K. Sagara~{\it et al.}, contribution to this conference.

\bibitem{deltuva:05b}
A. Deltuva, A.~C. Fonseca, A. Kievsky, S. Rosati, P.~U. Sauer, M. Viviani,
  Phys.~Rev.~C  71  (2005)  064003.

\bibitem{deltuva:05c}
A. Deltuva, A.~C. Fonseca, P.~U. Sauer, Phys.~Rev.~Lett. 95  (2005)  092301.

\bibitem{fonseca:fb18}
A. Deltuva and A.~C. Fonseca, contribution to this conference.

\end{thebibliography}

\end{document}